\documentclass[sigconf]{acmart}
\usepackage{multirow}
\usepackage{colortbl}
\usepackage[autostyle]{csquotes} 
\usepackage[skip=0pt]{caption}
\usepackage{hyperref}
\usepackage[
    type={CC},
    modifier={by},
    version={4.0},
    imagewidth=0.38\linewidth
]{doclicense}
\usepackage{wrapfig}
\AtBeginDocument{%
  \providecommand\BibTeX{{%
    \normalfont B\kern-0.5em{\scshape i\kern-0.25em b}\kern-0.8em\TeX}}}

\copyrightyear{2023}
\acmYear{2023}
\setcopyright{rightsretained}
\acmConference[CHCHI 2023]{Chinese CHI 2023}{November 13--16, 2023}{Denpasar, Bali, Indonesia}
\acmBooktitle{Chinese CHI 2023 (CHCHI 2023), November 13--16, 2023, Denpasar, Bali, Indonesia}\acmDOI{10.1145/3629606.3629646}
\acmISBN{979-8-4007-1645-4/23/11}

\begin{document}

\title{Decoding Fear: Exploring User Experiences in Virtual Reality Horror Games}


\author{He Zhang}
\email{hpz5211@psu.edu}
\affiliation{%
  \institution{The Future Laboratory, Tsinghua University}
  \streetaddress{No. 160 Chengfu Road}
  \city{Beijing}
  \country{China}}
\affiliation{%
  \institution{College of Information Sciences and Technology, Pennsylvania State University}
  \streetaddress{Westgate building}
  \city{University Park}
  \country{United States}}
\orcid{0000-0002-8169-1653}

\author{Xinyang Li}
\affiliation{%
  \institution{Academy of Arts \& Design, Tsinghua University}
  \city{Beijing}
  \country{China}}
  \orcid{0000-0003-4081-604X}

\author{Christine Qiu}
\affiliation{
  \institution{School of Electrical Engineering and Computer Science, KTH Royal Institute of Technology}
  \city{Stockholm}
  \country{Sweden}
}\orcid{0000-0002-9069-6851}
  
\author{Xinyi Fu}
\authornote{Corresponding author}
\affiliation{%
  \institution{The Future Laboratory, Tsinghua University}
  \streetaddress{No. 160 Chengfu Road}
  \city{Beijing}
  \country{China}}
\email{fuxy@mail.tsinghua.edu.cn}
\orcid{0000-0001-6927-0111}

\renewcommand{\shortauthors}{Zhang and Li, et al.}

\begin{abstract}
This preliminary study investigated user experiences in VR horror games, highlighting fear-triggering and gender-based differences in perception. By utilizing a scientifically validated and specially designed questionnaire, we successfully collected questionnaire data from 23 subjects for an early empirical study of fear induction in a virtual reality gaming environment. The early findings suggest that visual restrictions and ambient sound-enhanced realism may be more effective in intensifying the fear experience. Participants exhibited a tendency to avoid playing alone or during nighttime, underscoring the significant psychological impact of VR horror games. The study also revealed a distinct gender difference in fear perception, with female participants exhibiting a higher sensitivity to fear stimuli. However, the preference for different types of horror games was not solely dominated by males; it varied depending on factors such as the game's pace, its objectives, and the nature of the fear stimulant.
\end{abstract}



\maketitle

\section{Introduction}\label{sec:Introduction}
 Virtual reality (VR) has demonstrated immense potential across various fields, particularly in human-computer interaction (HCI)~\cite{kim2016virtual}. Over time, researchers and developers have focused their attention on improving the operability, accessibility, usability, and maintainability of VR systems. However, as VR devices already meet a certain level of usability and functionality, the value of enhancing the quality of VR applications based on design thinking becomes increasingly crucial. 

Emotions play an integral role in human society and have significantly impacted the progression of human-computer interaction technologies~\cite{picard1999affective}. Particularly, they hold substantial influence in human-centered design. Nonetheless, our grasp on human emotions, behaviors, and strategies in VR settings is still in need of deepening~\cite{7002681}.

Emotions, notably fear, are complex psychological phenomena involving cognitive and physiological aspects~\cite{solomonOpponentprocessTheoryAcquired1980, izard1991psychology}. Fear, a strong negative emotion, is triggered by perceived or actual danger and serves as a survival mechanism~\cite{lynchNothingFearAnalysis2015, clasen2011primal}. The Diagnostic and Statistical Manual of Mental Disorders (DSM-IV)~\cite{spitzer1994dsm} identifies five categories of fear-provoking factors, including animals, environment, blood/injections/injuries, situational factors, and other factors.

VR games, particularly horror games, have proven to be effective in eliciting fear responses from players, owing to the immersive and realistic experiences they offer~\cite{cantorFRIGHTREACTIONSMASS2008,slaterPlaceIllusionPlausibility2009,linFearVirtualReality2017,10.1145/3290607.3312832,lemmens2022fear,9597450}. Various game design techniques, including architectural spaces, movement restrictions, sound effects, and jump scares, are employed to induce fear~\cite{kalinowski2019silent,perron2004sign}. 

When fear is evoked, individuals employ various coping strategies, such as self-help strategies, approach (monitoring) strategies, and avoidance strategies~\cite{lynchNothingFearAnalysis2015,linFearVirtualReality2017}. Physiological responses, including changes in heart rate, facial expressions, and specific action tendencies, are common indicators of fear~\cite{kivikangas2018emotion}. 

Understanding fear stimuli is essential in studying user reflection in VR horror environments. Fears are usually classified as innate and acquired~\cite{agras1969epidemiology}, where innate fears are inherent and include~\cite{gibson1960visual}, for example, falling from a height, death or near death, loud noises~\cite{sbravatti2019acoustic}, high-altitude flying, etc. At the same time, experiences or memories mainly cause acquired fears, such as claustrophobia, darkness, dentists, socializing, snakes, bugs, thunderstorms, illness, loneliness, rejection, etc. Also, the Diagnostic and Statistical Manual of Mental Disorders (DSM-IV)~\cite{spitzer1994dsm} framework classifies fear-provoking factors into five categories: animals (e.g., dogs or spiders); environment (e.g., fire or floods); blood/injections/injuries (e.g., wounds or needles); situational factors (e.g., height, confined spaces or more specific spaces such as a doctor's office); and other factors (objectively harmless but disturbing stimuli such as distorted faces and loud noises). These fear factors mainly stimulate the amygdala, a region of the brain, to make people feel fearful~\cite{ledoux2012evolution}. In addition, the fear of paranormal phenomena, such as ghosts, is often used as a stimulus to appear in games or movies~\cite{perron2018world}. Overall, these stimuli have three core mechanisms in the VR environment, namely place illusion (PI, i.e., the strong illusion of being somewhere even though one knows one is not there), plausibility illusion (PSI, i.e., the illusion of being very real even though one inherently knows it is not real), and body ownership (i.e., the illusion of being the real self even though one is controlling an avatar). (i.e., although the person being controlled is an avatar, there is still the illusion of a real self-body)~\cite{slater2009place}. 

\section{User Experiment Design}
 The gaming processes include the specific process of experiencing the three games, the task settings, and the questionnaire collection.

\subsection{VR Horror Games Selection and Usage}  
In this study, we selected three VR horror games known to induce fear from the Steam platform: \textit{Richie's Plank Experience}~\cite{toast2016richie}, \textit{Phasmophobia}~\cite{phasmophobia}, and \textit{Emily Wants To Play}~\cite{hitchcock2015emily}. The selection criteria were as follows: a linear game script for controlling variables, game durations between 5-20 minutes to maintain emotional arousal, simple operation rules for easy learning, presence of elements to stimulate fear, and positive player ratings. 
\begin{figure*}[!ht]
  \centering
  \includegraphics[width=\textwidth]{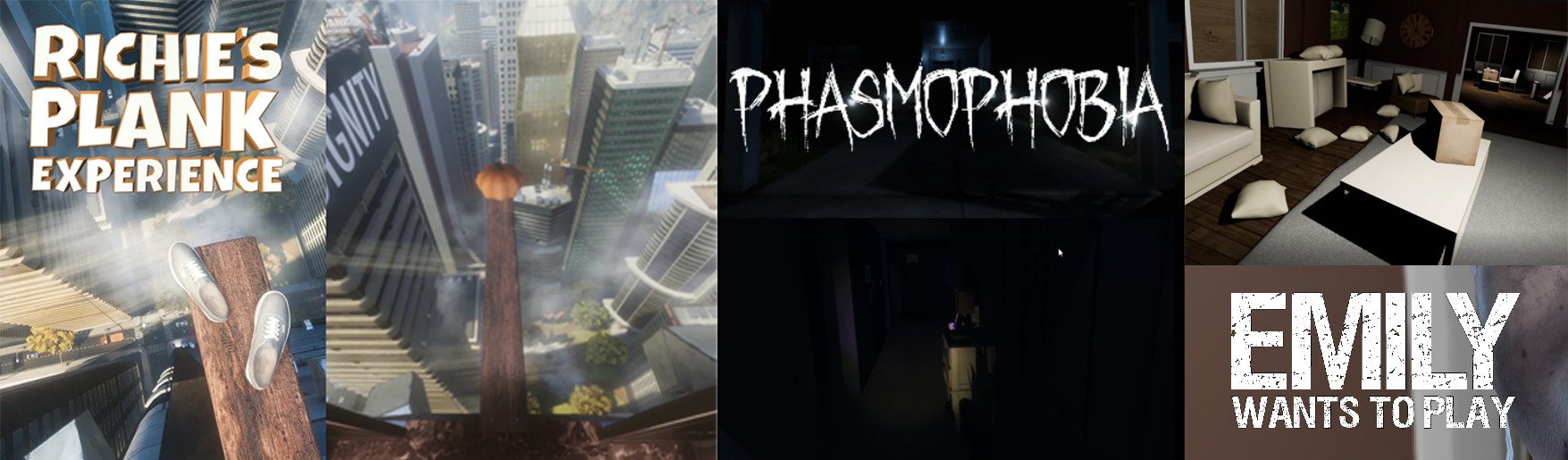}
  \caption{Examples of game posters and screenshots of the actual game. The first image on the left is the "Richie's Plank Experience" poster~\cite{toast2016richie}, the second image on the left is the "Richie's Plank Experience" in-game screenshot, the third image on the left is the "Phasmophobia" poster~\cite{phasmophobia}, the third image on the left is the "Phasmophobia" in-game screenshot, the first image on the right is the "Emily Wants To Play" in-game screenshot, and the first image on the right is the "Emily Wants To Play" poster~\cite{hitchcock2015emily}.}
  \label{fig:gamelist}
\end{figure*}

To reduce uncontrolled variables due to individual differences, we concealed or altered some actual goals, providing plausible but misleading objectives~\cite{kors2015foundation}. These allowed us to ensure the plot progression of games 2 and 3, independent of the player's actions. For instance, in Game 2, players were instructed to seek a nonexistent red bear toy, encouraging them to explore dark settings and trigger more fear-inducing events. In Game 1, no adjustments to gameplay were required as player actions did not alter the game progression.

\subsection{Participant}\label{subsec:data report of participant}
We recruited those who were interested in participating in the VR experiment as participants through social media groups and nearby universities in China, but for health and safety reasons, the eligible candidates for this experiment were controlled to be between 18-30 years old and self-reported to be in good physical health, mentally fit, and free of diseases or medical histories such as heart disease, visual problems, and vertigo. All participants were paid a certain amount of cash as thanks. All participants were given the option to stay after the experimental period for a free experience with VR games for additional time (with no special restrictions) without interfering with the experiment or the typical research environment. This experiment was processed under approval of university IRB. Informed consent was obtained from all participants.

A total of 23 participants' questionnaire data were successfully recorded, aged between 18 and 28 years old (median age = 21) and without significant congenital disorders. All of the participants are Chinese. All participants completed the pretest and posttest Panas-X scales~\cite{watson1994panas}, which is because the scale measures negative emotions and has been proven reliable in VR environment by numerous scientific studies~\cite{bagozzi1993examination,susindar2019feeling,drummond2021violent,newman2022use,steinhaeusser2022joyful}. Each participant who experienced a specific game completed the questionnaire for the corresponding game.

\section{Human Factors Analysis based on Experiments}
Gaining insights into participants' authentic reactions to the VR horror game experience is of paramount importance to all stakeholders. As an initial study, our study employed both validated scientific surveys and bespoke questionnaires. Our intent was to assemble quantifiable data from participants about the game's operability and its efficacy in evoking fear. This approach furnishes a wealth of insightful and actionable data that can significantly contribute to the trajectory of future research in this domain.

\subsection{Analysis and Preliminary Results}
We first explore the effects of VR horror games on human emotions and reactions through scales. We set up questionnaires before and after the experiment, and in each stage, all of the questionnaires asked participants to base their self-perceptions on what they were feeling at the time of completing the questionnaire. The purpose of the questionnaire was twofold: first, to determine roughly whether the chosen VR game evokes fear through the questionnaire statistics, and second, to explore possible future research questions. Here, we were more descriptive in our analysis of the questionnaire.

\subsubsection{Game Experience Evaluation Feedback Scale}
At the end of each game, we requested all participants to fill out a scale that was designed to be called the "Game Experience Evaluation Feedback Scale". This scale asked participants to rate for their feedback on the game they had just finished experiencing. The scale consisted of ten questions, each containing 7 rating scales (7-point Likert scale~\cite{joshi2015likert}). Participants could rate the questions on a scale from 1 to 7, where 1 is the lowest rating (completely negative) and 7 is the highest rating (completely positive). We first checked the reliability of the questionnaire using Spearman-Brown reliability coefficient (Equation~\ref{eqn:spearman-brown})~\cite{eisinga2013reliability} and theta reliability coefficient (Equation~\ref{eqn:theta})~\cite{zumbo2007ordinal}. 
\begin{equation}
\label{eqn:spearman-brown}
\rho_{xx} = 2r_{hh}/(1 + r_{hh})~,
\end{equation}
where $\rho_{xx}$ was the reliability estimates for the entire test, $r_{hh}$ was correlation coefficient of the two halves of the test scores. 
\begin{equation}
\label{eqn:theta}
\theta = N(1 - 1/\lambda)/N-1~,
\end{equation}
where N was the number of analysis items, $\lambda$ was root value of maximum characteristic.
The results in Table~\ref{tab:reliability statistics Spearman} and~\ref{tab:theta reliability coefficient} showed that the value of Spearman-Brown reliability coefficient(0.696) \textgreater~0.6 and the value of theta reliability coefficient(0.838) \textgreater~0.8 indicated that the scale data reliability quality was acceptable.

\begin{table}[ht]
\centering
\resizebox{\linewidth}{!}{%
\renewcommand{\arraystretch}{0.6} 
\begin{tabular}{lllll}
\toprule
\multicolumn{3}{l}{Reliability Statistics - Spearman-Brown coefficient}  \\ \midrule
Cronbach's Alpha &  & Part 1 & Value & 0.237 \\
 &  &  & N of Items & 5$^a$ \\
 &  & Part 2 & Value & 0.322 \\
 &  &  & N of Items & 5$^b$ \\
 &  & Total N of Items &  & 10 \\ 
Correlation Between Forms &  &  &  & 0.535 \\ 
Spearman-Brown Coefficient &  & Equal Length &  & 0.697 \\
 &  & Unequal Length &  & 0.697 \\ 
Guttman Split-Half Coefficient &  &  &  & 0.696 \\ 
$^a$The items are: Q1, Q2, Q3, Q4, Q5.\\
$^b$The items are: Q6, Q7, Q8, Q9, Q10 &  &  &  &  \\ \bottomrule
\end{tabular}}
\caption{Reliability Statistics - Spearman-Brown coefficient.}
\label{tab:reliability statistics Spearman}
\end{table}

\begin{figure*}[htbp]
  \centering
  \includegraphics[width=0.8\textwidth]{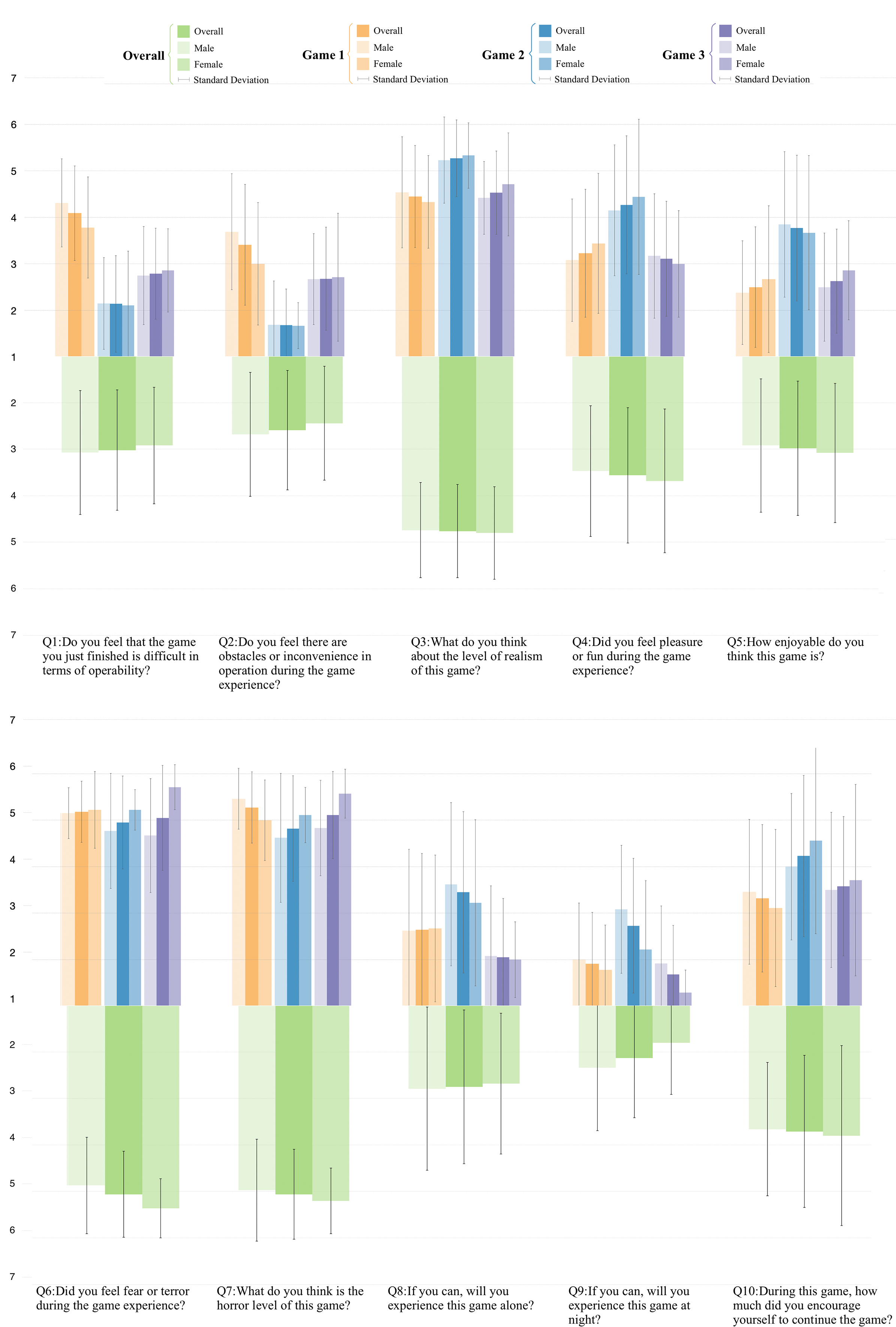}
  \caption{Game Experience Evaluation Feedback Scale. The statistics were formatted as mean (M) ± standard deviation (SD) in figure, i.e., M ± SD. Rounding to the 2 decimal places.}
  \label{fig.feedback scale}
\end{figure*}

The specific questions and results of the scale were shown in Figure~\ref{fig.feedback scale}, where the results included the average ratings of the participants collectively, the results of each game, and the gender-based groupings. 
The specific questions and results of the scale were shown in Figure~\ref{fig.feedback scale}, where the results included the average ratings of the participants collectively, the results of each game, and the gender-based groupings. 

Questions 1-3 and 5 were designed to verify that the game was appropriate for all experienced participants. Questions 4 and 6-10 were more specific to the topic of horror games. This scale's dataset contained 63 valid data due to the previously mentioned problems (severe 3D vertigo and too much fear), causing not all participants to complete all three games. The results showed that participants generally perceived the game operation as not difficult (Q1 and Q2 scored less than 4). Interestingly, from the results reported by the participants, women seemed to find the operation of VR games easier than men. At the same time, only 9(14.29\%) questionnaires reported the experience of the game as not realistic or immersive enough (Q3 scored less than 4). Analysis of the subgroups shows that Game 2 was reported to have a more authentic experience (5.27~$\pm$~0.83, mean score for Q3) compared to Game 1 (4.45~$\pm$~1.10) and Game 3 (4.53~$\pm$~0.90). Combined with the game content and the interviews mentioned in the previous section, participants of Game 2 seem to reflect an interesting observation that players are more likely to feel a sense of realism in games with visual limitations (darkness), realistic scenes, slow walking and accompanying ambient sounds (rain, footsteps, etc.). In the Q4, participants reported feeling less pleasure or happiness during the game experience (mean rating less than 4), which seems to be easily explained in conjunction with Q6. Since the emotions of pleasure, happiness, and fear, dread are opposites, inspiring fear and dread is more in line with the design purpose of horror games. After Pearson correlation coefficient analysis, Q4 and Q6 were found to have a significant negative correlation (see Table~\ref{tab:CorrelationQ4Q6}). In total, 53 reports indicated feeling very fearful and scared during the game experience (Q6 rating greater than 4), and of these, 2 reports gave an absolutely positive answer (rating 7). The results proved that the three games were successful for the activation of participants' terror emotions (overall scores greater than 5 for Q6 and Q7).
Combined with the actual data, female participants, on average, felt more fearful than male participants in all three games, especially in Game 3, where 100\% of female participants felt a higher level of fear rating (rating greater than or equal to 5). The results (Q8 and Q9) showed that participants were mainly reluctant to experience the three games in the experiment alone and would hardly choose to experience the games at night, which may reinforce that the games have a strong activating on the emotion of terror. Combined with the results of Q10, participants may have lower subjective fear feelings for Game 1 relative to Game 2 and Game 3. Females, overall, may be more sensitive to the horror elements in the game during play.

\subsubsection{Pre-test and Post-test PANAS-X}
\begin{table*}[ht]
\centering
\resizebox{\textwidth}{!}{%
\begin{tabular}{llr}
\toprule
Emotion Scales                &                &         \\
Emotional Groups by Positive or Negative& Emotional Groups by Types &Emotional Items\\ \hline
Basic Negative Emotion Scales(1.46~$\pm$~0.78, 1.79~$\pm$~0.87) & \textbf{Fear}(1.68~$\pm$~1.06, 2.78~$\pm$~1.10)           & \textbf{afraid}(1.96~$\pm$~1.15, 3.04~$\pm$~1.30), \textbf{scared}(1.61~$\pm$~1.12, 3.13~$\pm$~1.10), \\&&\textbf{frightened}(1.61~$\pm$~1.12, 3.83~$\pm$~1.03),  \textbf{nervous}(2.13~$\pm$~1.29, 2.09~$\pm$~1.12), \\&&\textbf{jittery}(1.35~$\pm$~0.65, 2.43~$\pm$~0.84), \textbf{shaky}(1.67~$\pm$~1.26, 2.17~$\pm$~1.23) \\
                              & Hostility(1.29~$\pm$~0.58, 1.43~$\pm$~0.78)      & angry(1.30~$\pm$~0.56, 1.35~$\pm$~0.65), hostile(1.30~$\pm$~0.56, 1.61~$\pm$~0.89), \\&&irritable(1.48~$\pm$~0.85, 1.43~$\pm$~0.66), scornful(1.13~$\pm$~0.34, 1.48~$\pm$~0.90), \\&&disgusted(1.17~$\pm$~0.49, 1.22~$\pm$~0.74), loathing(1.35~$\pm$~0.65, 1.48~$\pm$~0.85) \\
                              & Guilt(1.41~$\pm$~0.77, 1.37~$\pm$~0.81)          & guilty(1.39~$\pm$~0.66, 1.35~$\pm$~0.78), ashamed(1.35~$\pm$~0.78, 1.57~$\pm$~1.08), \\&&blameworthy(1.39~$\pm$~0.66, 1.30~$\pm$~0.76), angry at self(1.39~$\pm$~0.78, 1.39~$\pm$~0.78), \\&&disgusted with self(1.30~$\pm$~0.76, 1.13~$\pm$~0.63), \\&&dissatisfied with self(1.65~$\pm$~0.98, 1.48~$\pm$~0.85) \\
                              & Sadness(1.45~$\pm$~0.72, 1.56~$\pm$~0.79)        & sad(1.30~$\pm$~0.63, 1.57~$\pm$~0.73), blue(1.65~$\pm$~0.83, 1.70~$\pm$~0.82), \\&&downhearted(1.22~$\pm$~0.42, 1.78~$\pm$~1.04), alone(1.52~$\pm$~0.90, 1.30~$\pm$~0.70), \\&&lonely(1.57~$\pm$~0.84, 1.43~$\pm$~0.66)   \\ 
Basic Positive Emotion Scales(2.54~$\pm$~0.95, 2.33~$\pm$~1.11) & Joviality(2.80~$\pm$~0.96, 2.39~$\pm$~1.29)      & happy(2.78~$\pm$~1.00, 2.26~$\pm$~1.25), joyful(2.91~$\pm$~0.95, 2.48~$\pm$~1.27), \\&&delighted(2.65~$\pm$~1.07, 2.04~$\pm$~1.26), cheerful(2.74~$\pm$~0.81, 2.52~$\pm$~1.27), \\&&excited(2.78~$\pm$~0.85, 2.74~$\pm$~1.54), enthusiastic(2.70~$\pm$~0.88, 2.48~$\pm$~1.34), \\&&lively(2.83~$\pm$~1.03, 2.13~$\pm$~1.22), energetic(3.04~$\pm$~1.07, 2.43~$\pm$~1.20) \\
                              & Self-Assurance(2.34~$\pm$~0.97, 2.00~$\pm$~0.99) & proud(2.13~$\pm$~0.81, 1.74~$\pm$~0.96),  strong(2.35~$\pm$~1.07, 2.17~$\pm$~1.03), \\&&confident(2.74~$\pm$~0.92, 2.39~$\pm$~1.08), bold(1.78~$\pm$~0.85, 1.35~$\pm$~0.65), \\&&daring(2.65~$\pm$~1.07, 2.13~$\pm$~1.22), fearless(2.39~$\pm$~1.12, 2.22~$\pm$~1.00) \\
                              & Attentiveness(2.48~$\pm$~0.93, 2.61~$\pm$~1.04)  & alert(2.22~$\pm$~0.95, 3.13~$\pm$~0.76), attentive(2.91~$\pm$~1.15, 2.65~$\pm$~1.34), \\&&concentrating(2.78~$\pm$~0.80, 2.78~$\pm$~1.20), determined(2.00~$\pm$~0.80, 1.87~$\pm$~0.87) \\ 
Other Affective States(2.07~$\pm$~0.9825, 2.21~$\pm$~1.16)        & Shyness(1.69~$\pm$~0.95, 2.13~$\pm$~1.05)        & shy(1.61~$\pm$~0.99, 2.52~$\pm$~1.24), bashful(1.91~$\pm$~1.04, 1.39~$\pm$~0.66), \\&&sheepish(1.52~$\pm$~0.85, 1.91~$\pm$~1.04),  timid(1.70~$\pm$~1.15, 2.70~$\pm$~1.26)   \\
                              & Fatigue(1.90~$\pm$~0.91, 2.04~$\pm$~1.18)        & sleepy(1.78~$\pm$~0.95, 1.65~$\pm$~1.07), tired(2.43~$\pm$~0.95, 2.78~$\pm$~1.35), \\&&sluggish(1.65~$\pm$~0.83, 1.78~$\pm$~1.13),  drowsy(1.74~$\pm$~0.92, 1.96~$\pm$~1.15)\\
                              & Serenity(3.06~$\pm$~1.03, 2.10~$\pm$~1.18)       & calm(3.26~$\pm$~1.05, 2.22~$\pm$~1.28), relaxed(3.00~$\pm$~1.24, 2.00~$\pm$~1.13), \\&&at ease(2.91~$\pm$~0.79, 2.09~$\pm$~1.12)  \\
                              & Surprise(1.64~$\pm$~1.04, 2.55~$\pm$~1.23)       & amazed(1.57~$\pm$~0.99, 2.65~$\pm$~1.23), surprised(1.70~$\pm$~1.11, 2.39~$\pm$~1.20), \\&&astonished(1.65~$\pm$~1.03, 2.61~$\pm$~1.27) \\ \bottomrule
\end{tabular}%
}
\caption{Item Composition of the PANAS-X Scales and Participant Statistics. The pre-test and post-test average scores(M) with standard deviation(SD) for all participants are shown in parentheses after each item. Rounding to the 2 decimal places. Emotions that directly reflect fear are bolded in the table. An example of the format is "Item (pre-test M~$\pm$~SD, post-test M~$\pm$~SD)".}
\label{tab:panas-xstatistics}
\end{table*}

At the beginning and end of the experiment, participants were asked to fill out a PANAS-X each. The PANAS-X contains 60-item emotion descriptors with a five-point scale system, expressing participants' negative and positive emotions through a higher-order scale and responding to unique emotions through a lower-order scale~\cite{watson1994panas}. We refer to Watson and Clark's work~\cite{watson1994panas} and divided the scale into three subscales (Table~\ref{tab:panas-xstatistics} or Figure~\ref{fig.panas-xstatistics}), namely (1) Basic Negative Emotion Scales, which contains 4 emotional sets related to the emotions of fear, hostility, guilt, and sadness respectively; (2) Basic Positive Emotion Scales, which contains 3 emotional sets related to the emotions of joviality, self-assurance, attentiveness and (3) Other Affective States, which contains 4 emotional sets related to the emotions of shyness, fatigue, serenity, and surprise. Each type of emotional sets contains a number of unique emotional items. Cronbach's Alpha~\cite{tavakol2011making} was performed on the pre-test and post-test scales, and the results showed excellent reliability for both the pre-test($\alpha = 0.917$) and post-test($\alpha = 0.919$) scales. We conducted paired samples t-tests for the fearful emotion group on the Basic Negative Emotion Scale, and the results(Table~\ref{tab:pairedTest} and~\ref{tab:Paired Samples Correlations}) showed significant (p<0.05) changes in participants' ratings of fear-related emotions (scared, afraid, frightened, and jittery) before the start of the experiment and after the end of the experiment and significant (p<0.05) correlation in emotions(scared and afraid) between results of pre-test and post-test.  As reflected in the actual scores (Table~\ref{tab:pairedFear}), frightened emotion score increased by 2.22, jittery emotion increased by 1.08, scared emotion increased by 1.52, feared emotion increased by 1.08, and shaky emotion increased by 0.47. In summary, we believe that the chosen game has a good ability to stimulate participants' fear. In addition to this, our descriptive statistical analysis of the before and after results revealed some interesting phenomena, such as that after playing the fear game, (1) participants' sense of security decreased substantially, while (2) they were more alert than before the game (alert, surprise emotions went up, sleepy emotions went down), but led to other decreases in concentration and to feelings of exhaustion (tired, sluggish, drowsy, etc.).

The results of the user experiments showed that gender difference is a factor in the level of fear perception and that people adopt different ways of alleviating fear, all of which could be points of reference for future VR horror game research.


\section{Discussion and Future Work}
The potent induction of fear amongst participants underlines the efficacy of VR horror games in eliciting their intended emotional responses. Intriguingly, our findings suggest a noticeable gender difference in fear perception, with female participants exhibiting greater sensitivity to fear stimuli. However, preliminary results show that female players are likelier to have fun in slower-paced games. In addition, preliminary results suggest that females do not necessarily show more fear than males in all types of stimuli (Fig.~\ref{fig.feedback scale}. Q7 and Q10, also in Fig.~\ref{fig.panas-xstatistics-fear-gender}). In investigating innate fears, (Game 1) females seem to have a better tolerance level than males, but the results are reversed in darkness-related environments. These revelations signal exciting opportunities for future research to delve deeper into these gender-based discrepancies. Exploration into physiological responses, the impact of personality traits, or the influence of prior experiences on fear perception in VR games could broaden our understanding of this aspect.

The preference of players for games offering visual limitations and ambient sounds, perceived as more realistic, uncovers a promising pathway for future VR game design. Additional research could extend upon this premise, experimenting with diverse combinations of sensory stimuli and game mechanics to heighten the sense of realism and immersion. An enriched understanding of how individual elements contribute to presence can pave the way for the development of highly immersive and thrilling VR horror games.

Our discovery that participants were reluctant to partake in the games alone or at night intimates the profound psychological impact of these experiences. This realization has potential implications for designing VR horror games, particularly emphasizing user safety and well-being. Future explorations could incorporate multiplayer modes or safety mechanisms, such as tranquil intermissions or content filters, to modulate the intensity of fear stimuli. Further studies could examine how these adaptations influence players' fear responses and overall satisfaction.

The significant surge in fear-related emotions post-experiment, coupled with a substantial decrease in feelings of security, accentuates the profound impact of VR horror games on users' psychological states. As VR technology continues to advance, becoming increasingly immersive, it becomes crucial to understand and mitigate potential adverse effects on users' mental health. Future initiatives should explore strategies to provide players with post-game support to facilitate a return to their normal emotional states. This could include relaxation exercises or debriefing sessions.

Finally, in order to gain a more comprehensive understanding of the user's experience, future studies should consider incorporating semi-structured interviews. The quantitative data collected through our questionnaire sheds significant light on participant reactions and the elements contributing to fear induction. However, to gain a more nuanced understanding of these experiences, future research endeavors should consider the integration of semi-structured interviews. This methodology can add a rich, qualitative dimension to our data, providing detailed accounts of individual experiences, subjective emotions, and any new observations~\cite{qu2011qualitative}. The flexibility inherent to this approach allows researchers to delve deeper into specific questionnaire responses and further explore emergent themes or issues. For example, such interviews can probe into why certain game elements or environments elicit heightened fear responses in participants or why they may be reluctant to engage with the games alone or at night. By navigating these discussions, we can uncover the subjective insights and personal narratives that extend beyond the reach of quantitative data. 

Furthermore, semi-structured interviews could provide a platform to investigate broader topics~\cite{braun2012thematic}, such as the potential long-term psychological impacts of VR horror games or strategies players employ to cope with the fear stimuli. These qualitative insights could be instrumental in shaping future VR horror game design, ensuring they offer immersive and thrilling experiences and consideration of user safety and mental well-being. Although involving qualitative research might lead to a significant increase in data volume, with the improved performance of large-scale language models, using more automated tool~\cite{zhang2023qualigpt} or method~\cite{zhang2023redefining} to handle such data will be promising in the future.

Our preliminary results offer valuable insights that can guide the design of these future semi-structured interviews. Additionally, these results could be a meaningful reference for subsequent studies investigating fear induction in VR horror games. By combining these approaches, we aim to paint a more comprehensive picture of the VR horror gaming landscape and its psychological implications.


\begin{acks}
This work was supported by the National Natural Science Foundation of China (Grant No. 62172252)
\end{acks}

\bibliographystyle{ACM-Reference-Format}
\bibliography{sample-base}

\appendix
\section{APPENDICES}
\begin{table}[ht]
\centering
\resizebox{\linewidth}{!}{%
\begin{tabular}{cccc}
\hline
\multicolumn{2}{l}{Reliability Statistics} \\ \hline
Item && theta coefficient of item has been deleted & theta coefficient \\
Q1 && 0.846 & \multirow{10}{*}{0.838} \\
Q2 && 0.846 &  \\
Q3 && 0.848 &  \\
Q4 && 0.797 &  \\
Q5 && 0.804 &  \\
Q6 && 0.827 &  \\
Q7 && 0.824 &  \\
Q8 && 0.798 &  \\
Q9 && 0.803 &  \\
Q10 && 0.812 &  \\ \hline
\end{tabular}}
\caption{Reliability Statistics - theta reliability coefficient. Process in an online statistical tool, SPSSAU.}
\label{tab:theta reliability coefficient}
\end{table}
\begin{figure*}[htbp]
  \centering
  \includegraphics[width=\textwidth]{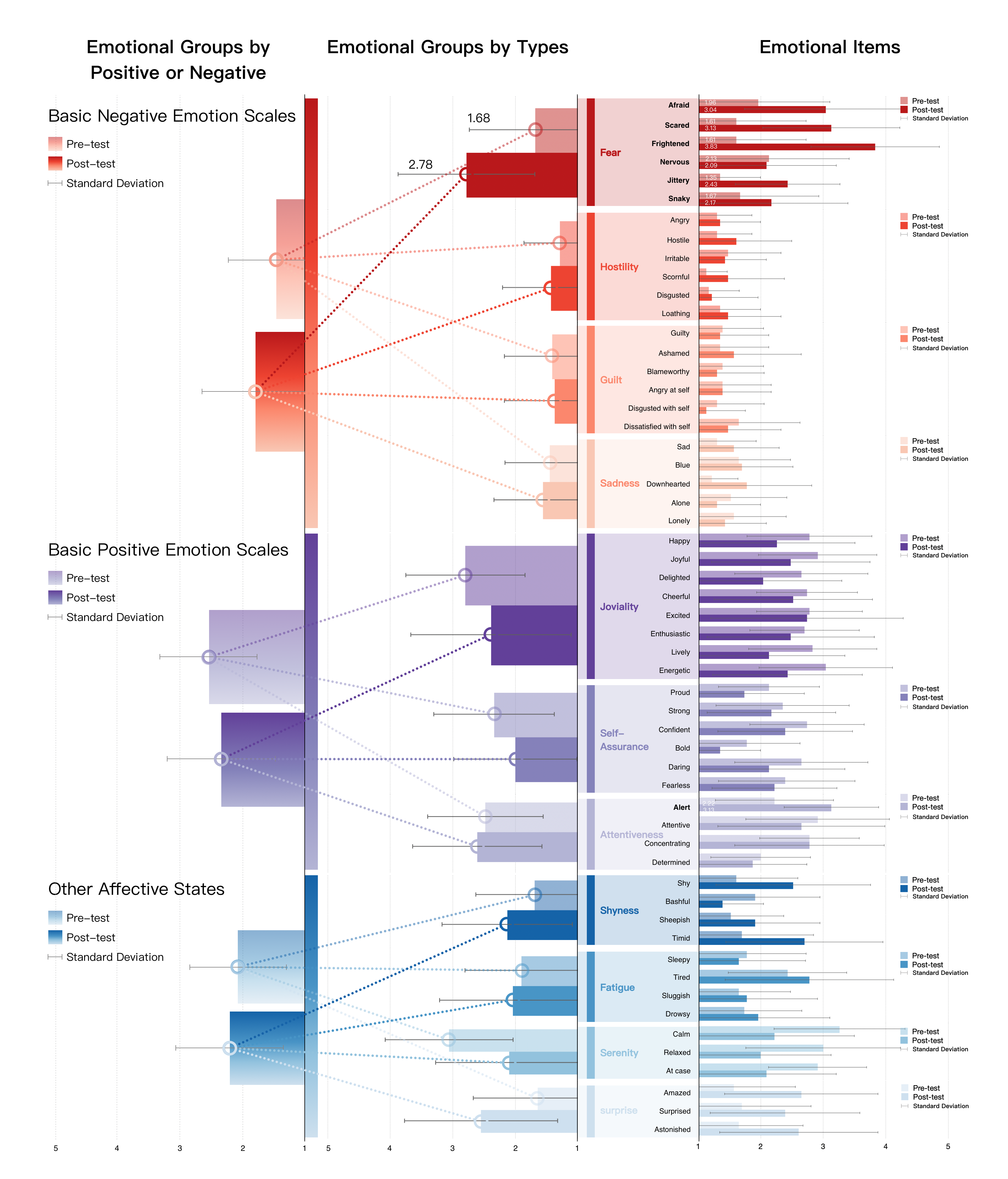}
  \caption{Item Composition of the PANAS-X Scales and Participant Statistics. The pre-test and post-test mean scores (M) and standard deviations (SD) of all participants are compared in this figure. Pre-test data for the same category are lighter in color compared to post-test data Rounding to the 2 decimal places. The original data are presented in Table~\ref{tab:panas-xstatistics}} 
  \label{fig.panas-xstatistics}
\end{figure*}
\begin{figure*}[htbp]
  \centering
  \includegraphics[width=\textwidth]{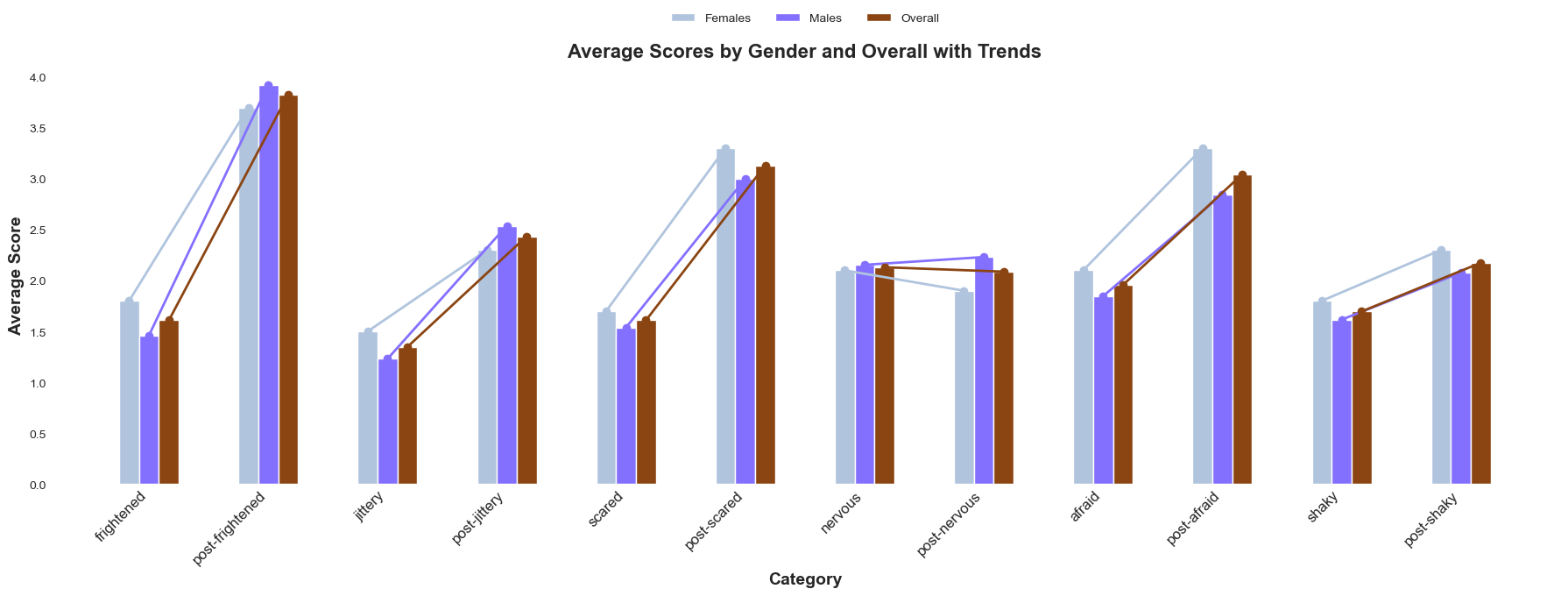}
  \caption{Item Composition of the PANAS-X Scales (Fear category) and Participant Statistics by genders. The pre-test and post-test mean scores (M) of all participants are compared in this figure.} 
  \label{fig.panas-xstatistics-fear-gender}
\end{figure*}

\begin{table}[htbp]
\centering
\resizebox{\linewidth}{!}{%
\begin{tabular}{llllll}
\toprule
\multicolumn{6}{l}{Paired Samples Statistics - Group of Fear Emotions} \\ \midrule
& & Mean & N & Std. Deviation & Std. Error Mean     \\
Pair 1 & frightened & 1.61 & 23 & 1.118 & .233 \\
 & post-frightened & 3.83 & 23 & 1.029 & .215 \\
Pair 2 & jittery & 1.35 & 23 & .647 & .135 \\
 & post-jittery & 2.43 & 23 & .843 & .176 \\
Pair 3 & scared & 1.61 & 23 & 1.118 & .233 \\
 & post-scared & 3.13 & 23 & 1.100 & .229 \\
Pair 4 & nervous & 2.13 & 23 & 1.290 & .269 \\
 & post-nervous & 2.09 & 23 & 1.125 & .235 \\
Pair 5 & afraid & 1.96 & 23 & 1.147 & .239 \\
 & post-afraid & 3.04 & 23 & 1.296 & .270 \\
Pair 6 & shaky & 1.70 & 23 & 1.259 & .263 \\
 & post-shaky & 2.17 & 23 & 1.230 & .257 \\ \bottomrule
\end{tabular}}
\caption{Paired comparisons of each fear-related emotion(Afraid, Scared, Frightened, Jittery, Nervous, and Shaky) before and after the experiment, based on the results of PANAS-X scale.}
\label{tab:pairedFear}
\end{table}

\begin{table*}[ht]
\centering
\resizebox{\linewidth}{!}{%
\begin{tabular}{lllllllllll}
\toprule
\multicolumn{9}{c}{Paired Differences} & \multicolumn{2}{c}{Significance} \\\hline
 &  &  &  &  & \multicolumn{2}{c}{\begin{tabular}[c]{@{}c@{}}95\% Confidence Interval\\ of the Difference\end{tabular}} &  &  &  &  \\
 &  & Mean & Std. Deviation & Std. Error Mean & Lower & Upper & t & df & One-Sided p & Two-Sided p \\
Pair 1 & frightened - post-frightened & -2.217 & 1.413 & .295 & -2.828 & -1.606 & -7.527 & 22 & \textless{}.001* & \textless{}.001* \\
Pair 2 & jittery - post-jittery & -1.087 & .996 & .208 & -1.518 & -.656 & -5.234 & 22 & \textless{}.001* & \textless{}.001* \\
Pair 3 & scared - post-scared & -1.522 & 1.163 & .242 & -2.025 & -1.019 & -6.277 & 22 & \textless{}.001* & \textless{}.001* \\
Pair 4 & nervous - post-nervous & .043 & 1.581 & .330 & -.640 & .727 & .132 & 22 & .448 & .896 \\
Pair 5 & afraid - post-afraid & -1.087 & 1.203 & .251 & -1.607 & -.567 & -4.334 & 22 & \textless{}.001* & \textless{}.001* \\
Pair 6 & shaky - post-shaky & -.478 & 1.442 & .301 & -1.102 & .145 & -1.591 & 22 & .063 & .126 \\ \bottomrule
\end{tabular}%
}
\caption{Paired samples t-test of fear-related emotions based on the PANAS-X. There was a significant difference (*significant at p<=0.05) between pre-test and post-test in the responses to PANAS-X.}
\label{tab:pairedTest}
\end{table*}

\begin{table*}[ht]
\centering
\resizebox{\linewidth}{!}{%
\begin{tabular}{llllll}
\toprule
Paired Samples Correlations - Group of Fear Emotions &  &  &  &  &  \\ \hline
 &  & \multicolumn{4}{c}{Significance} \\
 &  & N & Correlation & One-Sided p & Two-Sided p \\
Pair 1 & frightened \& post-frightened & 23 & .136 & .268 & .537 \\
Pair 2 & jittery \& post-jittery & 23 & .127 & .282 & .565 \\
Pair 3 & scared \& post-scared & 23 & .450 & .016* & .031* \\
Pair 4 & nervous \& post-nervous & 23 & .148 & .249 & .499 \\
Pair 5 & afraid \& post-afraid & 23 & .521 & .005* & .011* \\
Pair 6 & shaky \& post-shaky & 23 & .329 & .063 & .125 \\ \hline
\end{tabular}%
}
\caption{Correlation coefficients for paired samples (pre and post test) of fear-related emotions based on the PANAS-X. There was a significant correlation (*significant at p<=0.05) between pre-test and post-test in the responses to PANAS-X.}
\label{tab:Paired Samples Correlations}
\end{table*}

\begin{table*}[ht]
\centering
\resizebox{12cm}{!}{%
\begin{tabular}{llllll}
\toprule
\multicolumn{3}{l}{Correlations (Q4 and Q6 of Game Experience Evaluation Feedback Scale)}\\\midrule
   &                     & & Q4     & Q6     \\
Q4 & Pearson Correlation & & 1      & -.288* \\
   & Sig. (2-tailed)     & &        & .022   \\
   & N                   & & 63     & 63     \\
Q6 & Pearson Correlation & & -.288* & 1      \\
   & Sig. (2-tailed)     & & .022   &        \\
N  & N                   & & 63     & 63     \\ \midrule
\multicolumn{3}{l}{* Correlation is significant at the 0.05 level (2-tailed).} \\ \bottomrule
\end{tabular}}
\caption{Correlation Analysis of Question 4 and Question 6 of the Game Experience Evaluation Feedback Scale.}
\label{tab:CorrelationQ4Q6}
\end{table*}

\end{document}